\documentclass[a4paper]{jpconf}
\usepackage{graphicx}
\usepackage{cite}

\begin{document}

\title{N-body model of magnetic flux tubes reconnecting in the solar atmosphere}

\author{L Giovannelli$^1$, F Berrilli$^1$, D Del Moro$^1$, S Scardigli$^1$, G Consolini$^2$,
M Stangalini$^3$, F Giannattasio$^2$, A Caroli$^1$, F Pucci$^1$, V Penza$^1$}
\address{$^1$Universit\`a di Roma ``Tor Vergata'', Via della Ricerca Scientifica, I-00133 Roma, Italy}
\address{$^2$INAF - IAPS, Via del Fosso del Cavaliere 100, I-00133 Roma, Italy}
\address{$^3$INAF - OAR, Via Frascati 33, I-00040 Monte Porzio Catone (RM), Italy}

\begin{abstract}
The investigation of dynamics of the small scale magnetic field on the Sun photosphere is necessary to understand the physical processes occurring in the higher layers of solar atmosphere due to the magnetic coupling between the photosphere and the corona.
We present a simulation able to address these phenomena investigating the statistics of magnetic loops reconnections.
The simulation is based on N-body model approach and is divided in two computational layers. 
We simplify the convection problem, interpreting the larger convective scale, mesogranulation, as the result of the collective interaction of convective downflow of granular scale.
The N-body advection model is the base to generate a synthetic time series of nanoflares produced by interacting magnetic loops.
The reconnection of magnetic field lines is the result of the advection of the magnetic footpoints following the velocity field generated by the interacting downflows. 
The model gives a quantitative idea of how much energy is expected to be released by the reconfiguration of magnetic loops in the quiet Sun.
\end{abstract}

\section{Introduction}

One of the main goals in solar physics remains the detailed explanation of
how the plasma in the higher layers of the Sun has a stable temperature of
the order of $10^6$ K. Although several possible non-thermal mechanisms
have been proved to be able to transport energy from the photosphere,
a detailed explanation of the conversion from magnetic energy to thermal and particle energy
is still in need of an answer. Besides, a quantitatively and statistically strong observative evidence 
of these mechanisms is still lacking \cite{demoortel2015}. 
Magnetic reconnection events are considered to be one of the main energy source for the heating of
the outer layers of the solar atmosphere (chromosphere and corona).
Although both the numerical implementation of the magnetohydrodynamics (MHD) theory and the computational power have evolved
rapidly in the last years \cite{peter2015},
it remains prohibitive to compute full MHD 3D simulations
on large temporal and spatial scales and within an environment containing multiple evolving magnetic flux tubes.
In order to simulate a computational box with the side of more than $10$ Mm on time scales of several hours,
we propose a simulation which starts from simplified numerical models able to
mimic the behavior of the magnetic loops in the solar atmosphere
without needing to solve time consuming MHD equations.
Although not solving the proper physical equations of the system, this class of models can provide 
a synthetic, but realistic, time series of magnetic energy release due to the reconfigurations of magnetic loops.

\subsection{Simplified Reconnection Model}

In a simplified reconnection model presented in 2006 \cite{hughes2003},
multiple loops evolve in space and time in a 3D domain, hereafter Hughes Multiple Loops model (HML).

Magnetic loops are represented by semicircumferences on a plane perpendicular to the photospheric plane, and their footpoints, of equal intensity magnetic field and opposite polarity, are anchored on the photosphere.
The footpoints move in a random walk,
representing the motion of magnetic field advected by the plasma flows.
The magnetic energy stored in the loop is proportional to its length and increases as the two footpoints
increase their distance.
New magnetic loops are injected in the system in order to simulate the magnetic emergence process.
Reconnection can occur in the HML model when either two loops collide in 3D space, or two footpoints
annihilate. Reconnection is only allowed if it shortens the combined length of the two colliding loops,
thus resulting in a release of energy. The model also takes into consideration cascade events in which
one reconnection triggers others.

The resulting probability distribution of energy released during reconnections events is well described by
a power law with a slope $\alpha_E = 3.0\pm 0.2$.
This value is slightly high compared to the slope obtained by observational data in the UV band of low energy flares
$\alpha_E = 1.5 \div 2.6$ \cite{charbonneau2001}.
Another difference compared to the flare observations is the absence of any temporal correlation in the waiting time
among reconnection events, maybe due to the random nature of the motion of magnetic footpoints.

In order to improve this point and account for the correct spatial and temporal correlations of the
process underneath the magnetic footpoints motion we implemented a model similar to the HML.
The new model had to include a simplified advection model able to provide a more realistic motion
of the footpoints.

\subsection{N-body Advection Model}

A fluidodynamical simulation \cite{rast1998} has proved that
the profile of the vertical and horizontal velocity fields originated by a downflow is well
approximated by an exponential law. The N-body Advection (NA) model \cite{rast2003} starts from this
result, representing a downflow as a point generating an attractive
velocity field profile: $ v = V e^{-\frac{d}{\sigma}} $
at a distance $d$.
Every downflow is a sink in the plasma and is the source of an attractive velocity
field directed to the downflow, attracting all the other downflows. All downflows
are advectively transported according to the horizontal velocity field generated
by all the others downflows.

New downflows are added in random positions on the domain;
the intensity of the downflow at $d=0$, $V$,
is assigned from a normal distribution where $\sigma = 1$ and 
$\mu =1$, values are normalized to $1$ km/s.
The domain of the simulation is square with periodic boundary conditions (toroidal geometry).
The $V$ parameter has an exponential decay over time, characterized by a constant
decay $\tau$, which is related to the coherence time of granules.
If the intensity at zero distance $V$ goes below the value of $0.0001$ km/s, then the downflow is removed, 
representing an evanescent downflow.
The spatial scale of the simulation is obtained interpreting $2 \sigma$, the
typical influence scale of a downflow, as the granular scale.
The NA model then performs a check on the distance between the downflows, if it finds $d_{ij}<1$ the 
two downflows are merged. The new value for the $V$ parameter is the sum of the old ones: 
$V_k = V_i + V_j$.
Mass conservation is imposed forcing the conservation of the total flow $F = \sum_i V_i$
at every time step, adding downflows randomly distributed on the domain.
The space-time scales are set for the smaller granular-scales and they spontaneously
self-organize in the mesogranular scale,
constituted by more stable and intense downflows. The size of the mesogranular
scale is set by the proper values in the $\sigma-F$ parameter space \cite{bart2006}.
Once the right spatial scale for menogranulation is set,
the NA model reproduce the observed temporal
scale for the mesogranular convective structures \cite{delmoro2004,berrilli2004,berrilli2005}.
Thus it is possible to use a solar NA model to impose the correct spatial and temporal behavior
to the magnetic footpoints of an HML-like model.

\section{PASTA Model}

A model combining the NA and HML models together was proposed \cite{bart2006},
introducing two main changes with respect to the HML model:

\begin{itemize}
 \item the advective velocity field generated by the collective interaction of
the downflows is used to move the passive magnetic footpoints with the correct space and time scales.
 \item the periodic boundary conditions of the NA model were extended to the HML model in order
to be able to produce loops with typical scales larger than the computational box.
\end{itemize}
The model is able to reproduce the power law distribution for the flare released energy
over three decades ($10^{23} \div 10^{26}$ erg) with an index $\alpha_E = 2.4 \pm 0.1$,
to be consistent with the observations.
Furthermore a waiting time distribution is found, which is also in agreement with the observations,
following an exponential law with $\alpha =4.3\pm0.3$.

To further implement the model and calibrate the energetic output of the synthetic nanoflares,
other observational parameters have to be included in the simulation. 
\begin{figure*}[t]
\begin{center}
\includegraphics[scale=.60]{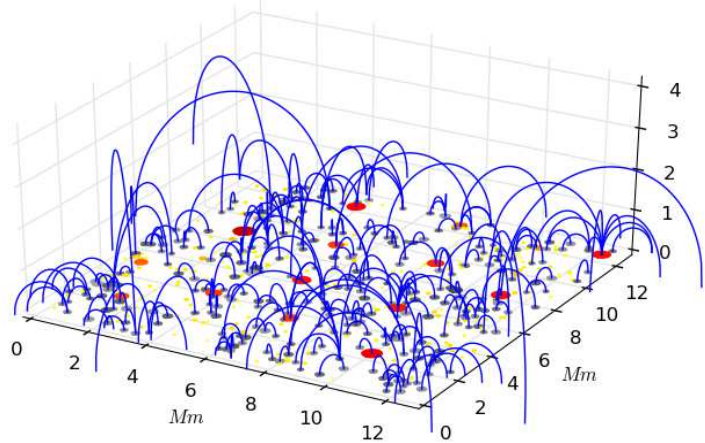}
\end{center}
\caption{3D representation of the PASTA simulation with an emergence
rate of $6.5~\cdot~10^{12} \;\mathrm{Mx} \;\mathrm{s}^{-1} \; \mathrm{Mm}^{-2}$. Yellow to red 
circles: downflows of increasing intensity. Grey circles: magnetic 
footpoints of positive (darker) and negative (lighter) polarity.
In both cases the circle radius is related to the intensity of the 
downflow or the magnetic field. 
Blue semicircumferences: magnetic loops.\label{fig1}}
\end{figure*}
We hereby present a new model, the Photospheric Advection Simulation at Tor vergata university ph.d. school of Astronomy
(PASTA), which derives directly from \cite{bart2006}.
We have rewritten completely the code in python, optimizing the mathematical routines
and adding three main features:

\begin{itemize}
 \item We have considered a distribution of intensity of the magnetic field. Within a magnetic loop
the two footpoints have the same strength and opposite polarity,
but the field intensity can vary from one loop to the other.
The intensity of the magnetic field is extracted from
the measured distribution of an exceptional 25 hours long dataset
of the HINODE SOT instrument \cite{giannattasio2013}.  
 \item Since the intensity of the magnetic field in two interacting loops is not the same,
we have introduced a new set of rules in case of interaction, described later in detail.
 \item The energy of the magnetic loop depends now on the length of the loop and on the intensity of the magnetic field.
\end{itemize}

\subsection{PASTA code description}

In our model, multiple loops evolve in space and time in a
3D domain as in \cite{hughes2003}.
The spatial resolution is set to $1$ pixel $= 50$ km.
The footpoints move accordingly to the velocity field generated
by the N-body model as in \cite{rast2003}, representing the motion of
magnetic field passively advected by the plasma.

New loops are added at a fixed rate.
When a new loop is added, the position of the 
first footpoint is random, the second footpoint of the same loop is put at distance 
$d_B$ in direction $\phi_B$. $d_B$ is extracted from a truncated normal distribution. 
The two linked footpoints have the same intensity for the magnetic field, but opposite polarity.
Magnetic footpoints are passively advected by the downflows, following the same 
spatial exponential law, and so behaving like corks.
In Fig. \ref{fig1} we show one step of the simulation
with the 3D loops connecting the magnetic footpoints and the underneath velocity field. 

We check for interaction events between footpoints, ruled by the interaction
distance set to the spatial resolution of the simulation, $1$ pixel, i.e. $50 km$:

\begin{itemize}
 \item CANCELLATION. If two footpoints of the same loop have a distance 
less than $3$ pixels, or a magnetic flux less than $10$ G the loop is canceled.

 \item AGGREGATION. If two footpoints of equal polarity have a distance less 
than $1$  pixel, the two footpoints are moved together in the same position, an average 
of the former ones. The footpoints are not merged, they remain distinct 
in order to prevent an arbitrary rule in case of reconnection from 
a footpoint with different loops departing from it. In this way we have the 
possibility to simulate a multiple loop footpoint avoiding ambiguity. 
The footpoints share the same position evolving together.

 \item ANNIHILATION. If two footpoints of different polarities of two different 
loops have a distance less than $10$ pixels, i.e. the granulation convective scale,
there is an annihilation event. 
These two footpoints are canceled
and a new loop is defined between the remaining other ones. 
We call loop A the loop with the more intense magnetic field and loop 
B the other. Loop A will interact with loop B as having a magnetic field with the 
same intensity of loop B. Loop A is schematized as made up of two loops, loop A', 
with the same intensity of loop B, and loop A'', with the remaining magnetic flux. 
In case of annihilation or reconnection events, only the A' and B loop interact, 
while the A'' loop remains untouched. At the end in case of an annihilation we will 
a have a C new loop and the A'' loop.

 \item RECONNECTION. If two loops, represented on the photosphere plane as segments, 
have an intersection on the periodic domain, named the X point, we measure the 
distance between the two points in the two loops corresponding to the X point. 
If this distance is less than $1$ pixel we perform a reconnection event, where we create two 
new loops connecting the positive footpoint of the first old loop with the negative 
footpoint of the second old loop and vice versa. The reconnection is performed 
only if the initial energy is greater than the final one; 
this guarantees a positive release of magnetic energy. 
This latter event is recorded as a synthetic nanoflare.
\end{itemize}

Multiple events at fixed time are 
allowed if a reconfigured loop has the possibility to reconnect with another loop 
after a first reconnection. In this way a flare resulting from a cascade event
of simultaneous reconnections is recorded. The possibility of one loop to cut other loops 
while reconfiguring, and in this way reconnecting with them, is not taken into 
consideration for simplicity.
We do not impose the conservation of the number of loops in the system, although the 
datasets show a steady number of loops during the simulation.\\
The energy of a loop is defined integrating the energy density $B^2/8\pi$
over the volume of the magnetic loop:
\begin{equation}
 E \mathrm{[erg]} = \frac{\pi}{16} \; d_{loop}\;(B \cdot B_r )^2
\end{equation}
where $B$ is the magnetic field intensity and $d_{loop}$ is the distance between the footpoints
on the photosphere and thus the diameter of the semicircumference.
The diameter of the magnetic flux tube, $2 B_r$, is set to the 
spatial resolution of the simulation, i.e. $50$ km.

\subsection{Exploring the Magnetic Emergence Rate Parameter}

We mainly focus our analysis on the exploration of the magnetic emergence rate parameter.
New measures of the emerging magnetic couples have shown that the previous
values known in the literature were underestimated at least by
an order of magnitude \cite{gonzalez2009, gonzalez2012}.
As we continue increasing spatial resolution and magnetic sensitivity, 
the small magnetic field contribution becomes very important 
in the emergence process of magnetic flux tubes.
We have investigated the relation between the simulated flare events 
and the emergence process, trying to estimate the energy released for 
the observed emergence rate and understand if it is sufficient to 
explain the coronal heating.

Six simulations are computed, with the same dynamical advection parameters,
and increasing magnetic emergence rate, from $7.8 \cdot 10^{11} \; \mathrm{Mx} \; \mathrm{s}^{-1} \; \mathrm{Mm}^{-2}$
to $2.6 \cdot 10^{13} \; \mathrm{Mx} \; \mathrm{s}^{-1} \; \mathrm{Mm}^{-2}$, doubling the magnetic emergence rate
from one simulation run to the other. The aim is to find out the
relation between magnetic emergence rate and the probability distribution of the energy released by the 
synthetic nanoflares.

The domain is a square matrix with a $256$ pixels side. The total intensity of the downflows
is set to $1000$.
The spatial scale is set by the $\sigma = 10$ parameter,
thus having $20 \; \mathrm{pixels} = 1 \; \mathrm{Mm}$. The temporal scale is set by the parameter 
$\tau = 10$ (simulation time unit), interpreted as the typical coherence time of a granule ($10$ minutes).
The six runs have a goal to simulate $600$ minutes.

\section{Results}

\begin{figure*}[t]
\begin{center}
\includegraphics[scale=.4]{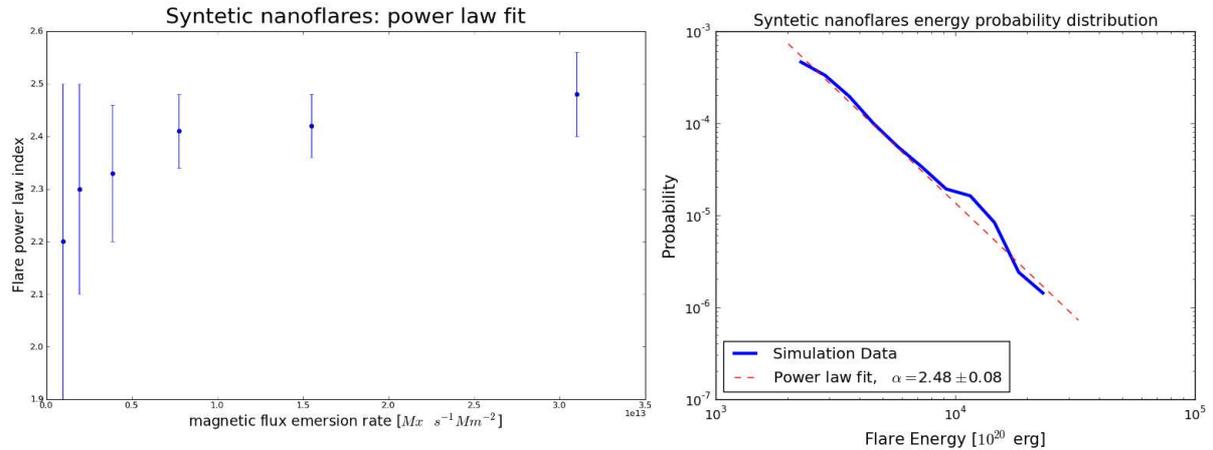}
\end{center}
\caption{Left: Power law index of the energy probability distribution for synthetic nanoflares
as a function of the magnetic flux emergence rate.
Right: example of the power law fit for the nanoflare energy PDF in the case of
the highest magnetic emergence rate.\label{fig3}}
\end{figure*}

We obtain from each simulation a synthetic nanoflare time sequence and
perform a statistical analysis on the released magnetic energy
probability distribution function (PDF).
PDFs are well described by a power law and the fit is obtained using the
method described in \cite{powerlaw}. \\
As it is clear from Fig. \ref{fig3} the index reaches a stable value for a sufficiently high emergence
rate of the order of $\sim 0.7 \; \mathrm{Mx} \; \mathrm{s}^{-1} \;\mathrm{Mm}^{-2}$.\\
As evident from Fig. \ref{fig3}, the simulation runs relative to the lowest magnetic emergence rate
have larger error bars for their power law index.
This is an effect of the poorer statistics of flare events in these three runs, due to the lower
emergence rate, which gives noisier PDFs.
Indeed, the probability to have a reconnection event is a function of the density of the loops,
which is not set by any conservation rules, but reaches a steady value non-trivially
determined by the magnetic emergence rate parameter.
Although the power law index values show an increasing trend as a function of the magnetic emergence rate,
taking into account the associated error bars, we can state that all six simulation runs have a
power law index compatible with $\alpha_E \simeq 2.4$ (see Fig. \ref{fig3}).



We use the synthetic nanoflare statistics
to understand if it is compatible with a plausible mechanism to 
heat the solar corona. We find that the computed power law index 
is compatible with the theoretical formulation by Parker,
which requires $\alpha > 2$,
for a  nanoflare energy distribution able to sustain a stable hot corona 
\cite{parker1988,vekstein2000,aschwanden2000}.

The computed power law index $\alpha \simeq 2.4$ is also compatible
with the index obtained from observations.
Solar flares show a similar behavior with a power law
index reported $1.5 < \alpha < 2.6$ \cite{charbonneau2001} and evaluated
in a reduced range $2.1 < \alpha < 2.6$ using geometrical assumptions
\cite{mcintosh2001}. Thus our model has $\alpha$ in
the range reported by the observers, although our energy range 
is between $2 \cdot 10^{20} \; \mathrm{erg} \; \div \; 3 \cdot 10^{24} \; \mathrm{erg}$,
while the observed energy range is extended at most to the
microflare range (see for example the reported power law index
$2.3 < \alpha < 2.6$
for the flare in the range
$ 10^{25} \; erg \; \div \; 3 \cdot 10^{26} \; erg$ from UV observations
\cite{krucker1998}).

\section{Summary}

The simulation performed in our study combines a model able to mimic the
correct spatial and temporal scales of granulation and mesogranulation 
with a simplified description of the reconnection events between
magnetic coronal loops.\\ 
The model is able to reproduce the observed dynamical properties of
small scale magnetic elements and produce a synthetic nanoflare time sequence
over $600$ minutes, on a computational domain equivalent to a square
with a $12.8$ Mm side of the solar phostosphere.\\
We have explored the magnetic emergence rate parameter of the simulation,
introducing a new observational constraint.
The nanoflare PDF is well fitted by a power law with an index compatible with
$\alpha \simeq 2.4$, which is in agreement with flare observations and compatible with
the requirement stated by Parker ($\alpha > 2$) for a nanoflare heated solar corona.

\end{document}